\documentclass[aps,prb,twocolumn,preprintnumbers,amsmath,amssymb]{revtex4}
\usepackage{graphicx}
\usepackage{xcolor}
\begin{document}
\title{Magnetoresistance of a bulk sample of FeSi.}

\author{A.~E.~Petrova}
\affiliation{P.~N.~Lebedev Physical Institute, Leninsky pr., 53, 119991 Moscow, Russia}
\author{S.~Yu.~Gavrilkin}
\affiliation{P.~N.~Lebedev Physical Institute, Leninsky pr., 53, 119991 Moscow, Russia}
\author{S.~S.~Khasanov}
\affiliation{Institute of Solid State Physics RAS, Chernogolovka, Moscow District,142432 Russia}
\author{V.~A.~Stepanov}
\affiliation{P.~N.~Lebedev Physical Institute, Leninsky pr., 53, 119991 Moscow, Russia}
\author{S.~M.~Stishov}
\email{stishovsm@lebedev.ru}
\affiliation{P. N. Lebedev Physical Institute, Leninsky pr., 53, 119991 Moscow, Russia}

\begin{abstract}
The magnetoresistance, MR  of a well-characterized bulk FeSi sample was studied. It is shown that after a chaotic behavior at temperatures below 6 K, the magnetoresistance of FeSi becomes regular functions of temperature and magnetic field. The observations  suggest that the mean free path of carriers defines along with the unknown negative component the negative values of  magnetoresistance of FeSi  then approaching  zero values at high temperatures. 
\end{abstract}
\maketitle

\section{Introduction}
The unusual physical properties of iron monosilicide, FeSi, have been attracting much attention for years. Indeed, FeSi being a metal at temperatures about 100-150~K, becomes a narrow gap semiconductor as temperature decreasing. That is accompanied by puzzling behavior of magnetic susceptibility, heat capacity, resistivity and etc. This situation was analyzed in a number of publications (see for instances~\cite{Jac,Hunt,Fu,Man,Pas,Ish,Ar,Tom}). In result FeSi was characterized as a strongly correlated insulator. Nevertheless, the physics of FeSi appears to bring more surprise. The softening of the phonon spectrum was discovered upon metallization in FeSi~\cite{Delaire}. The topological Weyl points were found at studying phonons by the inelastic X-ray scattering in FeSi~\cite{Miao}. The topological features of the electron spectra in FeSi were recently studied~\cite{Changdar}. Authors of Ref.~\cite{Changdar} claim metallic state of FeSi outside of the temperature range 75-143~K.  Recently, the low temperature metallic conductivity was discovered in FeSi~\cite{Yu}, indicating conductive surface state as it occurs in a topological insulator. Then comparing properties of the surface conductive FeSi and SmB$_{6}$ authors~\cite{Bre} concluded that the surface conductive FeSi is a new d-electron Kondo insulator. Noting that most of arguments in Ref.~\cite{Bre} based on magnetoresistance measurements, we decided to take a careful look at the magnetoresistance of the bulk sample of FeSi.

\section{Experimental} 
To prepare the sample we used the same batch of FeSi, grown by Czochralski technique, as in our former experiments (see for instance the review~\cite{SMS}). So, this material was rather completely characterized. However, we report here some new data adding new features to the description of our sample of FeSi. 
The chemical composition of the sample determined by the electron-probe microanalysis shows some excess of Fe content (Average data in atomic percents:  Fe 50.555; Si 49.045). The lattice parameter obtained from the Rietveld analysis of the powder X-ray diffraction appears to be equal 4.4860(1)~\AA. The overall mosaicity as follows from a rocking curve using (400) reflection is about  0.05$^{0}$. The sample selected for magnetoresistance measurements has dimensions $\approx 2.5 \times 0.9 \times 0.6~ mm^{3}$. Resistance was measured in a direction of [100] corresponding to the longest size of the sample.       

\begin{figure}[htb]
\includegraphics[width=80mm]{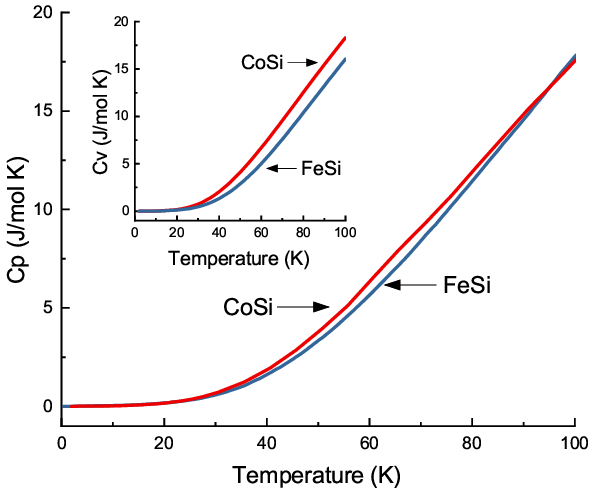}
\caption{\label{fig1} (Color online) Heat capacity $C_{p}$ of FeSi (current measurement) and CoSi~\cite{SMS}. In the inset one can see theoretically calculated phonon heat capacities of FeSi and CoSi~\cite{SMS}.}
\end{figure}

\begin{figure}[htb]
\includegraphics[width=80mm]{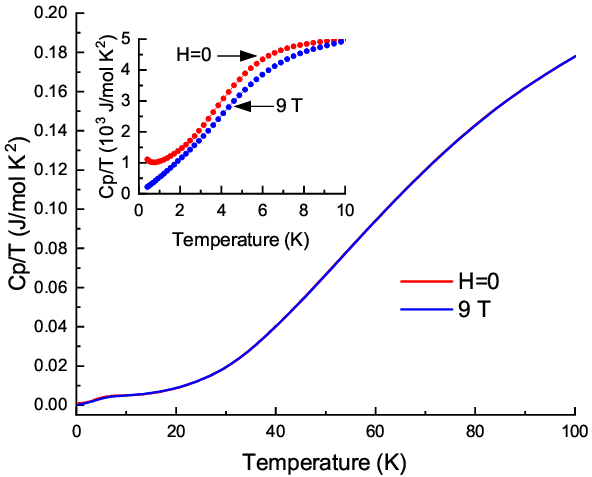}
\caption{\label{fig2} (Color online) $C_{p}/T$ of FeSi at zero magnetic field and 9~T.}
\end{figure}
\begin{figure}[htb]
\includegraphics[width=60mm]{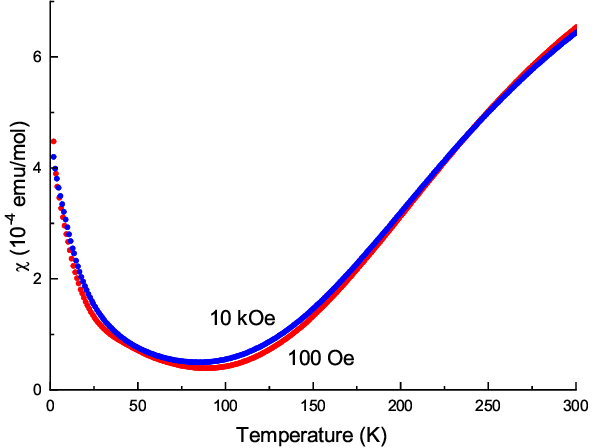}
\caption{\label{fig3} (Color online) The magnetic susceptibility of FeSi as function of temperature.}
\end{figure}

Heat capacity of the sample was measured making use the Quantum Design physical property measurement system with the heat capacity module and the He-3 refrigerator. The resistivity was measured with the standard four terminals scheme using gold wires glued to the sample by the silver paste as electrical contacts. Magnetic susceptibilities were measured with a Quantum Design magnetic properties measurement system.

The obtained heat capacity data are presented in Fig.~\ref{fig1} and Fig.~\ref{fig2} as $C_p/T$ verus $T$. As is seen the $C_p/T$ curve demonstrates apart from well-known hump at about 7~K some sort of a hook like feature near 1 K, which disappears in strong magnetic field. 

\begin{figure}[htb]
\includegraphics[width=80mm]{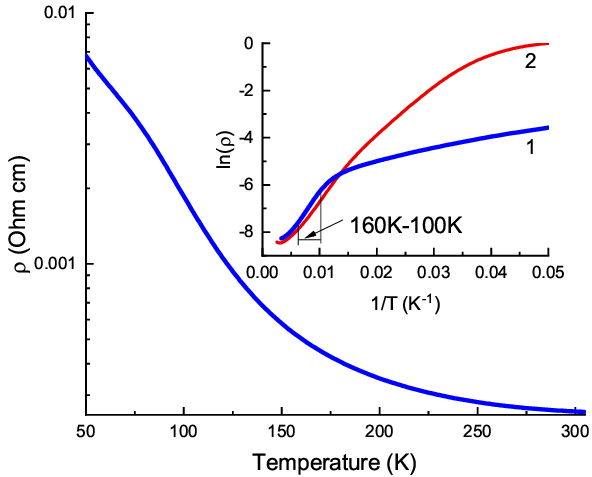}
\caption{\label{fig4} (Color online) Resistivity $\rho$ of FeSi as function of temperature T. In the inset two lines demonstrate behavior of resistivity of samples 1 and 2 in coordinates $ln(\rho)$ and T$^{-1}$. 1-bulk FeSi, 2-thread-shape FeSi~\cite{Yu}.}
\end{figure}

The magnetic susceptibility $\chi$ in Fig.~\ref{fig3} shows typical anomalous features of FeSi as was observed in a number of works. In particular, the low temperature increasing part of the $\chi(T)$ is often treated as a paramagnetic contribution attributed to residual disorder~\cite{Jac}.
The resistivity of FeSi (Fig.~\ref{fig4}) demonstrates a complex behavior, which includes different regimes (see also Table~\ref{table1}).

\begin{table}
\caption{\label{table1} Resistivity of the FeSi sample at selected temperatures.}

\begin{tabular}{|c|c|}
\hline 
$T$, K & $\rho$, Ohm cm \\ 
\hline 
1.83	 & 3.50\\ 
\hline 
77.05 & 	3.62x10$^{-3}$  \\ 
\hline 
300.02 & 2.58x10$^{-4}$\\ 
\hline 
\end{tabular} 
\end{table}

Figure~\ref{fig4} (see line 1 in the inset) shows that  in the narrow temperature region of 100–150~K the resistivity of our bulk FeSi can be described by a standard activation formula with an energy gap $E_g$ of 0.06~eV or 700~K. Line 2 in the inset is the resistivity curve of thread-shape FeSi, which as it claimed in Ref.~\cite{Yu} reveals a second gap in the range 30-54 K. Nothing like that is seen in bulk FeSi. A great difference in the sample characteristics is obvious.

\begin{figure}[htb]
\includegraphics[width=60mm]{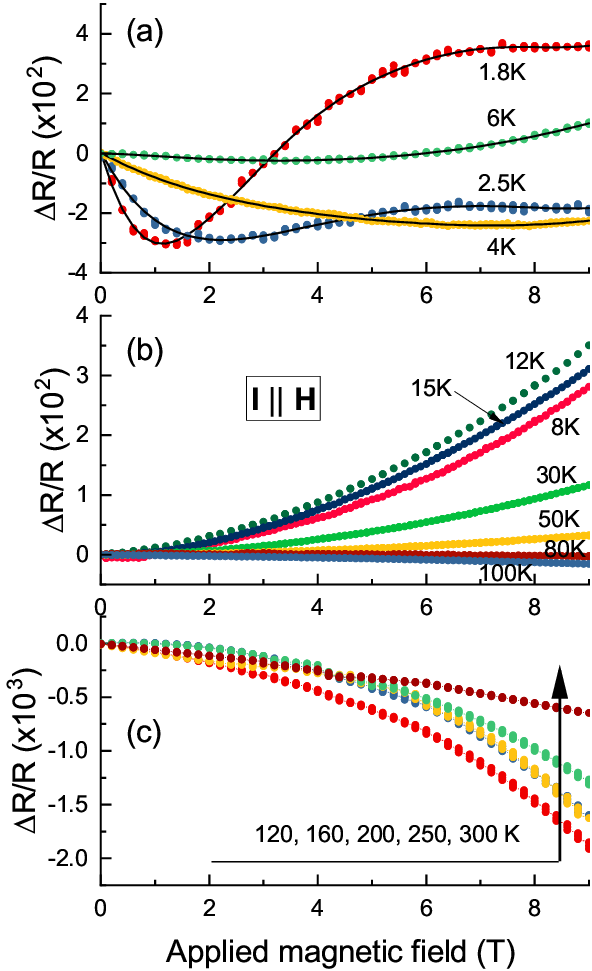}
\caption{\label{fig5} (Color online) Magnetoresistance of FeSi as functions of temperature and longitudinal magnetic field. Note change the scale $\Delta R/R$ in Panel C.  }
\end{figure}

\begin{figure}[htb]
\includegraphics[width=60mm]{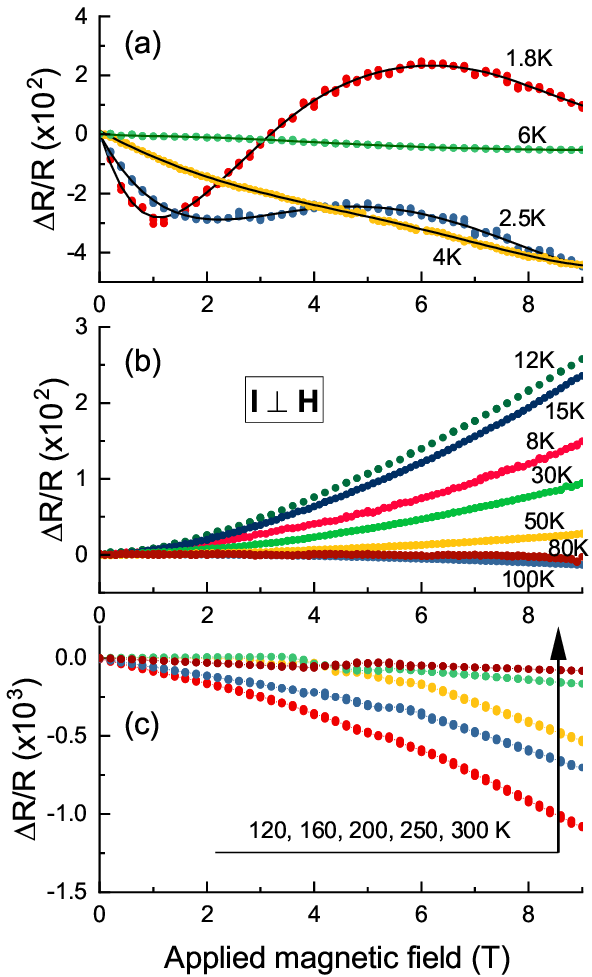}
\caption{\label{fig6} (Color online) Magnetoresistance of FeSi as functions of temperature and transverse magnetic field. Note change the scale $\Delta R/R$ in Panel C. }
\end{figure}

Measurements of magnetoresistance were carried out in two geometries with current parallel $I\| H$ and perpendicular $I\bot H$ to a magnetic field. At measurements the sample was cooled from 300 K to 1.8 K by steps (see Fig.~\ref{fig5}, \ref{fig6}. At each step magnetic field was applied from 0 to 9 T and the resistance measured in the configuration $I\bot H$ then measurement were continued in the configuration $I\| H$ on releasing of magnetic field from 9 T to 0. 
Results of magnetoresistance measurements are depicted in Fig.~\ref{fig5},~\ref{fig6}. They are distinctly can be distinguished in two groups differed by the temperature of measurements. On the other hand, one cannot see any significant difference in results obtained in two different geometries. Reversing the direction of the magnetic field at the resistance measurements of FeSi indicates a lack of significant Hall components in the current data.
	
\begin{figure}[htb]
\includegraphics[width=80mm]{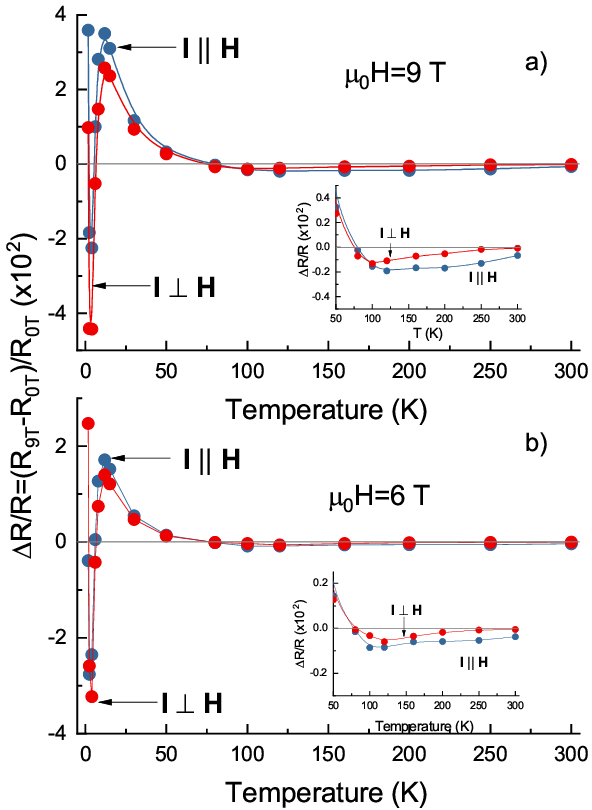}
\caption{\label{fig7} (Color online)  Temperature dependence of magnetoresistance of FeSi at magnetic fields 6 and 9~T.  }
\end{figure}
\begin{figure}[htb]
\includegraphics[width=60mm]{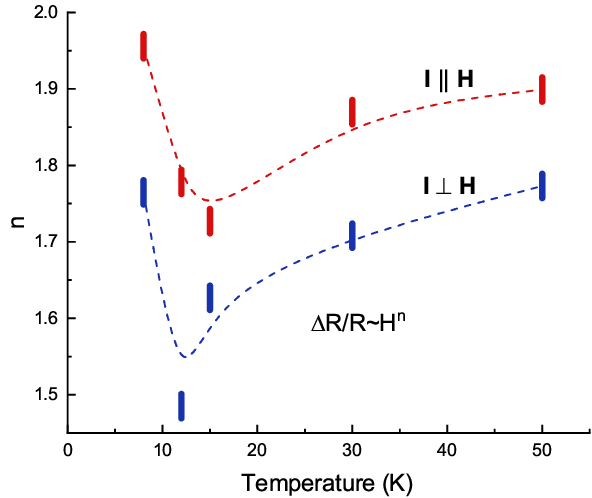}
\caption{\label{fig8} (Color online)  Values of exponent $n$ of expression $\Delta R/R  \sim H^{n}$, describing magnetoresistance curves in Fig.\ref{fig5} and Fig.\ref{fig6} at $T > 6K$.}
\end{figure}

\begin{figure}[h!] 
\includegraphics[width=80mm]{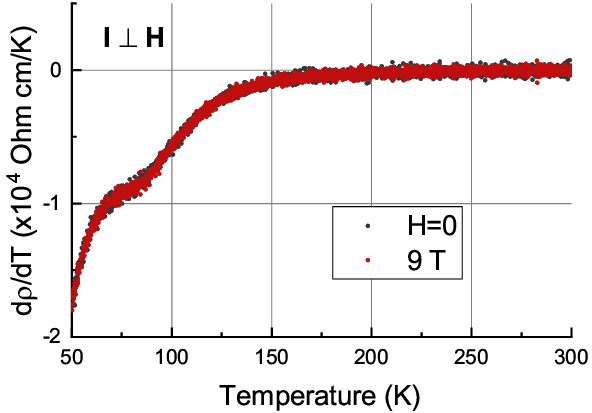}
\caption{\label{fig9} (Color online) The derivative of resistivity of FeSi at 0~T and 9~T and $I\perp H$. The same is seen in the parallel orientation.  Note that FeSi is on the verge of metallic behavior at 180-200~K. } 
\end{figure}
\begin{figure}[h!] 
\includegraphics[width=80mm]{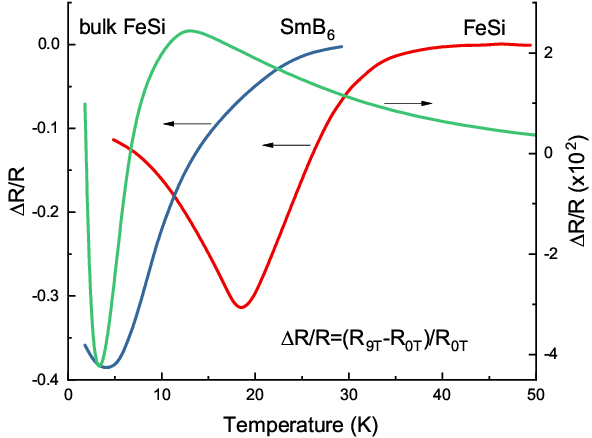}
\caption{\label{fig10} (Color online) Comparison of the magnetoresistance MR of FeSi and SmB$_{6}$ as a function of temperature at a magnetic field B=9~T after~\cite{Bre}. Current bulk FeSi data are also placed in the plot. Note that the magnetoresistance of bulk FeSi is smaller than MR from ~\cite{Bre} by the order of magnitude.}
\end{figure}

\section{Results and discussion}
It is accepted that at studying phonon properties of the B20 compounds CoSi may serve as a standard insulating nonmagnetic material with B20 crystal structure (see for instance~\cite{Jac}). Fig.~\ref{fig1} shows the experimental heat capacity curves of FeSi and CoSi in the 100-degree range~\cite{SMS}. The slight difference is obvious and actually corresponds to the theoretical calculations to 
about 60~K. But then the heat capacity of FeSi starts to grow fast which probably indicates an onset of metallization~\cite{Delaire}. 

We also mentioned before the hook like feature in $C_p/T$ at about 1~K. This feature was observed also in Ref.~\cite{Hunt}. Authors~\cite{Hunt} ascribed the rise of $C_p/T$ at low temperature to the Anderson localization. Note that magnetic field eliminates this specific of $C_p/T$  (see Fig.~\ref{fig2}).

Coming back to the magnetoresistance data (Fig.~\ref{fig5},~\ref{fig6}) we see (1) that they don't drastically sensitive to the measurement geometry, (2) they can be distinguished in three groups by temperature of measurements: low $T<6$ K, and high regions $6<T<100$ and $100<T<300$ K. That was not seen in some previous measurements of magnetoresistance in FeSi~\cite{Pas,Lis,Oh}. As a difference to the low $T$ group, the high $T$ groups behaves in some regular way as functions of magnetic field and temperature. As seen in Fig.~\ref{fig7} the magnetoresistance of high $T$ group passes a maximum at about 12~K and gradually decreases approaching  slightly negative values and then zero.

An extremum around 12 K is also seen in Fig.~\ref{fig8} demonstrating an dependence of exponent $n$ of expression $\Delta R/R  \sim H^{n}$ on temperature. It is seen that at temperature of 8~K the "longitudinal" isotermal exponent $n$ close to canonical value of 2. 
Despite the similarity of magnetoresistance of FeSi for perpendicular and parallel orientation of magnetic field and current (see Fig.~\ref{fig5} and Fig.~\ref{fig6}) it becomes anisotropic at magnetic fields higher than 3-4~T. 
At this discussion we are probably missing an important point which is the current jetting~\cite{Pip}. The latter effect might strongly influence the longitudinal magnetoresistance measurements. Hence, the corresponding data should be treated with much care because we could not control a situation in this respect.

We are not going here to analyze the magnetoresistance behavior of FeSi at T$\leqslant$ 6~K. This specific region, containing partly negative MR isoterms, is characterized by an interplay of localized and free carriers and possible occurring the quantum interference~\cite{Alt, Abrik}(see also~\cite{Pas}). So the each curve $\rho (M)$ would  require a special approach.

Turning to the higher temperature region at T$>6$~K one can see general decreasing the magnetoresistance of FeSi with temperature, which could be explained at the first sight by increasing the carrier concentration upon temperature induced metallization and  decreasing the mean free path of carriers. Yet, as shown by the maximum of MR at 12 K (Fig.~\ref{fig7}) it is not so simple. To describe the situation one may propose two components of magnetoresistance, positive and negative and with latter one starts rising around 12 K, as indicated in Fig.~\ref{fig7}. 
The results of the corresponding measurements, shown in Fig.~\ref{fig5}, \ref{fig6} and \ref{fig7}, probably support this point. As we see the magnetoresistance of FeSi becomes slightly negative at temperatures higher than  $\cong 60 K$. Symptomatic that the anisotropy in magnetoresistance of FeSi is not completely disappeared (Fig.~\ref{fig5},\ref{fig6}).  It is instructive to compare Figs.~\ref{fig4}, \ref{fig7}. As is seen the magnetoresistance of FeSi and greatly decreased in temperature range corresponding a semiconducting behavior but very close to transition to a metallic state~\cite{Sal}. Note that Fig.~\ref{fig9} shows also an existence of two branches in the curve $d\rho/dT$, which possibly connected to almost sudden increasing a number of carriers in the conductivity band.  On the other hand the plateau of $d\rho/dT$ in the range ~150-200 K of Fig.~\ref{fig9} indicate that the carrier concentration possibly reached its maximum and the mean free path, which probably became small enough,  defines along with the unknown negative component the negative values of  magnetoresistance of FeSi at high temperatures (see also~\cite{Sal}). Is a negative component of MR in FeSi connected to its chiral nature remains to be seen at future studies.

Now we turn our attention to Fig.~\ref{fig10}, which depicts figures illustrating of MR at magnetic field 9~T for the thread-shaped sample of FeSi and Kondo insulator SmB$_{6}$~\cite{Bre}, supposedly proving  their close similarity.  On the other hand the corresponding dependence for the bulk sample of FeSi from the current study reveals more in common with SmB$_{6}$ than the thread-shaped sample of FeSi though the magnetoresistance of bulk FeSi is smaller by the order of magnitude. Actually the minima in MR of SmB$_{6}$ and bulk FeSi are results of small mismatch of two high resistivity curves at zero magnetic field and 9~T, whereas the minimum in MR of thread-shaped FeSi connected to forming conductive surface at 19~K. 

\section{Conclusion}

The magnetoresistance of a well-characterized bulk FeSi sample was studied. The magnetoresistance data (Fig.~\ref{fig4},~\ref{fig5}) (1) don’t drastically sensitive to the measurement geometry, (2) they can be distinguished in two groups by temperature of measurements: low $T<6$ K, High $T>6$ K.  As a difference to the low $T$ group, the high $T$ group behaves in some regular way as functions of magnetic field and temperature. The magnetoresistance of high $T$ group passes a maximum at about 12~K and gradually decreases approaching  slightly negative values and then zero. Thus the observations  suggest that the mean free path of carriers defines along with the unknown negative component the negative values of  magnetoresistance of FeSi at high temperatures.  Is a negative component of MR in FeSi connected to its chiral nature remains to be seen at future studies.

Finally we should say that we couldn't see a second energy gap in the resistivity curve in our bulk sample of FeSi (Fig.~\ref{fig4}) as it claimed for the thread-shaped FeSi~\cite{Yu,Bre}. Moreover, a similarity between the temperature dependencies of magnetoresistance of the thread-shaped FeSi and Kondo insulator SmB$_{6}$~\cite{Bre} is questionable. Perhaps, the thread-shaped FeSi is an absolutely unique material and not necessary to put a known tag on it.

\end{document}